\documentclass[twocolumn,tighten]{aastex631}

\usepackage{amsmath}

\usepackage{multirow}
\usepackage{array}
\usepackage{booktabs}
\newcommand\DistZw{160}
\newcommand\RedshZw{0.036}
\newcommand\ScaleZw{310}
\newcommand\DistNGCthree{40.4}
\newcommand\RedshNGCthree{0.009}
\newcommand\ScaleNGCthree{78}
\newcommand\DistNGCseven{70.6}
\newcommand\RedshNGCseven{0.016}
\newcommand\ScaleNGCseven{137}
\newcommand\DistVV{84.4}
\newcommand\RedshVV{0.020}
\newcommand\ScaleVV{164}
\newcommand\SpecNum{267}
\newcommand\SpecnumZw{24}
\newcommand\SpecnumNGCthreeN{66}
\newcommand\SpecnumNGCthreeS{62}
\newcommand\SpecnumNGCseven{61}
\newcommand\SpecnumVV{54}
\newcommand\CRDS{11.17.16}
\newcommand\Pipeline{1.12.5}

\newcommand\SsilMin{-2.77}
\newcommand\SsilMax{0.23}
\newcommand\SsilMed{-0.43}
\newcommand\SFRmin{0.1}
\newcommand\SFRMax{3.1}
\newcommand\SFRMed{0.6}

\newcommand\AvSsilZw{-0.87}
\newcommand\AvSsilNGCseven{-0.19}
\newcommand\AvSsilNGCthirtyN{-0.28}
\newcommand\AvSsilNGCthirtyS{-0.87}
\newcommand\AvSsilVV{-1.13}
\newcommand\PAHelevenLowSilMed{-3.6}
\newcommand\PAHelevenLowSilIQRlow{-5.6}
\newcommand\PAHelevenLowSilIQRhigh{-1.6}

\usepackage{CJKutf8}

\begin{document}
\begin{CJK*}{UTF8}{bsmi}

\title{A Spectroscopically Calibrated Prescription for Extracting PAH Flux from JWST MIRI Imaging}

\author[0009-0001-6065-0414]{Grant P. Donnelly}
\affiliation{Ritter Astrophysical Research Center, University of Toledo, Toledo, OH 43606, USA}

\author[0000-0001-8490-6632]{Thomas S.-Y. Lai (賴劭愉)}
\affiliation{IPAC, California Institute of Technology, 1200 E. California Blvd., Pasadena, CA 91125, USA}

\author[0000-0003-3498-2973]{Lee Armus}
\affiliation{IPAC, California Institute of Technology, 1200 E. California Blvd., Pasadena, CA 91125, USA}

\author[0000-0003-0699-6083]{Tanio D\'iaz-Santos}
\affiliation{School of Sciences, European University Cyprus, Diogenes street, Engomi, 1516 Nicosia, Cyprus}
\affiliation{Institute of Astrophysics, Foundation for Research and Technology-Hellas (FORTH), Heraklion, 70013, Greece}

\author[0000-0003-3917-6460]{Kirsten L. Larson}
\affiliation{AURA for the European Space Agency (ESA), Space Telescope Science Institute, 3700 San Martin Drive, Baltimore, MD 21218, USA}

\author[0000-0003-0057-8892]{Loreto Barcos-Mu\~noz}
\affiliation{National Radio Astronomy Observatory, 520 Edgemont Rd, Charlottesville, VA, 22903, USA}
\affiliation{Department of Astronomy, University of Virginia, 530 McCormick Road, Charlottesville, VA 22903, USA}

\author[0000-0002-6570-9446]{Marina Bianchin}
\affiliation{Department of Physics and Astronomy, 4129 Frederick Reines Hall, University of California, Irvine, CA 92697, USA}

\author[0000-0002-4375-254X]{Thomas Bohn}
\affiliation{Hiroshima Astrophysical Science Center, Hiroshima University, 1-3-1 Kagamiyama, Higashi-Hiroshima, Hiroshima 739-8526, Japan}

\author[0000-0002-5666-7782]{Torsten B\"oker}
\affiliation{European Space Agency, Space Telescope Science Institute, Baltimore, MD 21218, USA}

\author[0009-0003-4835-2435]{Victorine A. Buiten}
\affiliation{Leiden Observatory, Leiden University, PO Box 9513, 2300 RA Leiden, The Netherlands}

\author[0000-0002-2688-1956]{Vassilis Charmandaris}
\affiliation{Department of Physics, University of Crete, Heraklion, 71003, Greece}
\affiliation{Institute of Astrophysics, Foundation for Research and Technology-Hellas (FORTH), Heraklion, 70013, Greece}
\affiliation{School of Sciences, European University Cyprus, Diogenes street, Engomi, 1516 Nicosia, Cyprus}

\author[0000-0003-2638-1334]{Aaron S. Evans}
\affiliation{National Radio Astronomy Observatory, 520 Edgemont Rd, Charlottesville, VA, 22903, USA}
\affiliation{Department of Astronomy, University of Virginia, 530 McCormick Road, Charlottesville, VA 22903, USA}

\author[0000-0001-6028-8059]{Justin Howell}
\affiliation{IPAC, California Institute of Technology, 1200 E. California Blvd., Pasadena, CA 91125}

\author[0000-0003-4268-0393]{Hanae Inami}
\affiliation{Hiroshima Astrophysical Science Center, Hiroshima University, 1-3-1 Kagamiyama, Higashi-Hiroshima, Hiroshima 739-8526, Japan}

\author[0000-0002-2603-2639]{Darshan Kakkad}
\affiliation{Centre for Astrophysics Research, University of Hertfordshire, College Lane, Hatfield AL10 9AB, UK}

\author[0000-0003-4023-8657]{Laura Lenki\'c}
\affiliation{IPAC, California Institute of Technology, 1200 E. California Blvd., Pasadena, CA 91125, USA}

\author[0000-0002-1000-6081]{Sean T. Linden} 
\affiliation{Steward Observatory, University of Arizona, 933 N Cherry Avenue, Tucson, AZ 85721, USA}

\author[0000-0003-0532-6213]{Cristina M. Lofaro}
\affiliation{Dipartimento di Fisica e Astronomia, Università di Padova, Vicolo dell’Osservatorio 3, 35122 Padova, Italy}
\affiliation{Institute of Astrophysics, Foundation for Research and Technology-Hellas (FORTH), Heraklion, 70013, Greece}

\author[0000-0001-6919-1237]{Matthew A. Malkan}
\affiliation{Department of Physics \& Astronomy, 430 Portola Plaza, University of California, Los Angeles, CA 90095, USA}

\author[0000-0001-7421-2944]{Anne M. Medling}
\affiliation{Department of Physics \& Astronomy and Ritter Astrophysical Research Center, University of Toledo, Toledo, OH 43606,USA}

\author[0000-0003-3474-1125]{George C. Privon}
\affiliation{National Radio Astronomy Observatory, 520 Edgemont Rd, Charlottesville, VA, 22903, USA}
\affiliation{Department of Astronomy, University of Virginia, 530 McCormick Road, Charlottesville, VA 22903, USA}
\affiliation{Department of Astronomy, University of Florida, P.O. Box 112055, Gainesville, FL 32611, USA}

\author[0000-0001-5231-2645]{Claudio Ricci}
\affiliation{Instituto de Estudios Astrof\'isicos, Facultad de Ingenier\'ia y Ciencias, Universidad Diego Portales, Avenida Ejercito Libertador 441, Santiago, Chile}
\affiliation{Kavli Institute for Astronomy and Astrophysics, Peking University, Beijing 100871, Peopleʼs Republic of China}

\author[0000-0003-1545-5078]{J.D.T. Smith}
\affiliation{Ritter Astrophysical Research Center, University of Toledo, Toledo, OH 43606, USA}

\author[0000-0002-3139-3041]{Yiqing Song}
\affiliation{European Southern Observatory, Alonso de Córdova, 3107, Vitacura, Santiago, 763-0355, Chile}
\affiliation{Joint ALMA Observatory, Alonso de Córdova, 3107, Vitacura, Santiago, 763-0355, Chile}

\author[0000-0002-2596-8531]{Sabrina Stierwalt}
\affiliation{Physics Department, 1600 Campus Road, Occidental College, Los Angeles, CA 90041, USA}

\author[0000-0001-5434-5942]{Paul P. van der Werf}
\affiliation{Leiden Observatory, Leiden University, PO Box 9513, 2300 RA Leiden, The Netherlands}

\author[0000-0002-1912-0024]{Vivian U}
\affiliation{Department of Physics and Astronomy, 4129 Frederick Reines Hall, University of California, Irvine, CA 92697, USA}

\begin{abstract}

We introduce a prescription for estimating the flux of the 7.7\,\micron\ and 11.3\,\micron\ polycyclic aromatic hydrocarbon (PAH) features from broadband JWST/MIRI images. Probing PAH flux with MIRI imaging data has advantages in field of view, spatial resolution, and sensitivity compared with MIRI spectral maps, but comparisons with spectra are needed to calibrate these flux estimations over a wide variety of environments. For \SpecNum\ MIRI/MRS spectra from independent regions in the four luminous infrared galaxies (LIRGs) in the Great Observatories All-sky LIRG Survey (GOALS) early release science program, we derive synthetic filter photometry and directly compare estimated PAH fluxes to those measured from detailed spectral fits. We find that for probing PAH 7.7\,\micron, the best combination of filters is F560W, F770W, and either F1500W or F2100W, and the best for PAH 11.3\,\micron\ is F560W, F1000W, F1130W, and F1500W. The prescription with these combinations yields predicted flux densities that typically agree with values from spectral decomposition within $\sim 7\%$ and $\sim 5\%$ for PAH 7.7 and 11.3\,\micron, respectively.

\end{abstract}

\section{Introduction}

The mid-infrared (MIR) radiation from the smallest dust grains in the interstellar medium can be used as a probe of the physical conditions within galaxies. Known as polycyclic aromatic hydrocarbons (PAHs), these grains reprocess the local interstellar radiation field, with the bulk of their spectral emission taking the form of prominent features at 3.3, 6.2, 7.7, 8.6, 11.3, 12.6, and 17\,\micron\ \citep{leger_puget_1984, allamandola_1985}. As PAH material preferentially absorbs ultraviolet radiation \citep{allamandola_1989, draine_2001} and it is abundant at the edges of molecular clouds in photo-dissociation regions (PDRs), PAH emission is often used as an infrared tracer of the star-formation rate in galaxies \citep{peters_2004, shipley_2016, lai_2020}. The relative strengths of the PAH spectral features are influenced both by physical properties of the grains such as their size and ionization state, as well as by the hardness and intensity of the radiation field \citep[e.g.][]{draine_2021}. This allows the observed ratios of these features to inform models of dust growth, destruction and photoionization \citep{narayanan_23, matsumoto_2024}.

The most accurate way to probe PAH emission is by using spectroscopy. As PAH spectral features exhibit wide ``wings", knowledge of the underlying continuum across a broad wavelength range is required to accurately extract the flux of PAH features \citep{Uchida_2000, smith_07, marshall_2007, Xie_2018}. Further, spectroscopy is required to accurately separate the fluxes of overlapping PAH features that arise from distinct vibrational modes in PAH molecules --- and thereby indicate different physical conditions --- in addition to separating the emission from overlapping atomic fine structure and recombination lines. Attenuation can add additional complexity to the MIR spectrum. In particular, the absorption feature attributed to silicate grains and centered at 9.7\,\micron\ can significantly alter the observed flux from the PAH 11.3\,\micron\ feature \citep{lai_2024}.

The more spatial coverage a PAH emission map has for a given object, the better it can capture how PAH characteristics change with environment. This is clear from studies of Milky Way regions \citep[e.g.][]{PDRs4All}, and of large sections of other galaxies \citep{sandstrom_2012,donnelly_2024,whitcomb_2024}. Further, high spatial resolution is necessary to explore key questions related to the survival or excitation of PAHs at varying distances to high-energy sources such as an active galactic nucleus (AGN) \citep[e.g.][]{diaz-santos_2010, jensen_17} or PDRs \cite[e.g.][]{egorov_2024, pedrini_2024}. The Spitzer Space Telescope could efficiently create spectral maps over whole galaxies \citep{kennicutt_2003, haan_2011}, but only at a spatial resolution of typically a few hundred parsecs for nearby galaxies. The James Webb Space Telescope (JWST) offers an order of magnitude improvement over Spitzer in spatial resolution and nearly two orders of magnitude in sensitivity for studying PAHs, but the relatively small fields of view (FOVs) of the NIRSpec and MIRI integral field units (IFUs) require a substantial amount of observing time to cover wide areas of nearby galaxies.

At the cost of spectral information, the imaging modules of NIRCam and MIRI onboard JWST offer a much larger FOV (over 100 times larger for MIRI) and better sensitivity to faint emission compared with the IFUs, while retaining high spatial resolution. These factors make it desirable to use imaging data to investigate PAH emission by photometric proxy over much larger areas and down to lower surface brightness than is often practical with spectral mapping, and work using JWST images of nearby galaxies has demonstrated this to be effective \citep{chastenet_2023, chastenet_2023_b}. JWST is particularly suited to image PAH emission as it offers multiple bands centered on PAH features for low-redshift targets: NIRCam/F335M, MIRI/F770W, and MIRI/F1130W. 

However, it is also necessary to estimate the continuum contribution in these bands to accurately estimate PAH flux. Efforts to extract the PAH flux from the NIRCam/F335M band benefit from the flanking F300M and F360M bands that can be used to estimate the local continuum \citep{inami_2018, lai_2020, sandstrom_23, gregg_2024, bolatto_2024}, but there are now MIRI bands specifically designed for continuum subtraction. 

Recently, \cite{chown_prescription} showed that PAH flux can be accurately estimated with MIRI photometry based on images and IFU spectroscopy of the Orion Bar on scales of a few parsecs. In this work, we provide a spectroscopically calibrated method for estimating the observed flux of the PAH 7.7 and PAH 11.3\,\micron\ features using only MIRI photometry down to $\sim 7\%$ and $\sim 5\%$, respectively. We test and calibrate this prescription on a diverse set of \SpecNum\ regions on spatial scales of $\sim80-300\,\mathrm{pc}$ with varying spectral slope, star-formation rate, and degree of obscuration in the four luminous infrared galaxies (LIRGs) from the Great Observatories All-sky LIRG Survey (GOALS) early release science (ERS) sample. We provide details of the calibration for various combinations of MIRI bands that are available.

\begin{figure*}[htb]
    \includegraphics[width=1\textwidth]{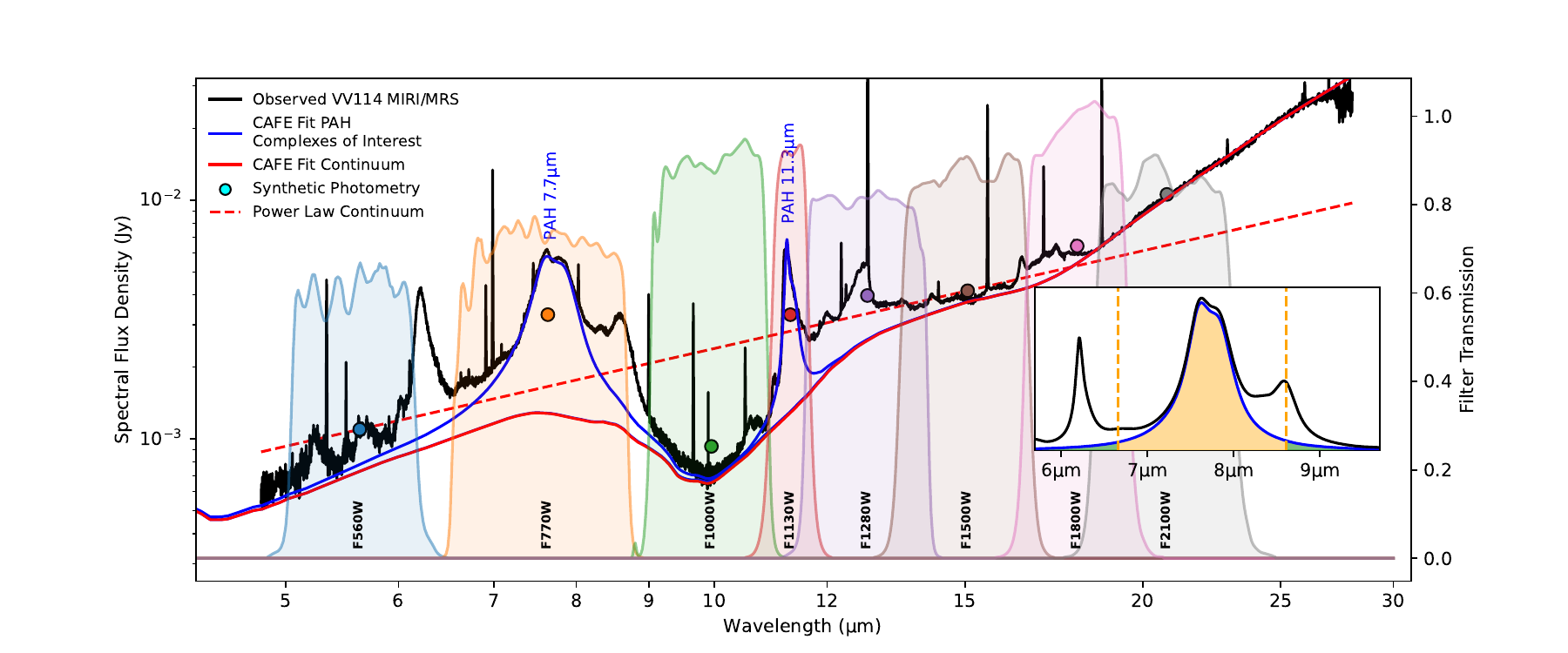}
    \caption{An example rest-frame MIRI/MRS spectrum from a region within VV114. The maximum flux values have been truncated for clarity. The blue line shows the \texttt{CAFE}-fit contribution from the two PAH features discussed in this work, and the solid red line is the \texttt{CAFE}-fit continuum. Filled circles indicate the value of the synthetic photometry for each MIRI filter, shown as shaded areas. The dashed red line shows an example power-law continuum (see \S~\ref{sec:powerlawcont}) anchored at F560W and F1500W. In the inset, we show a section of the \texttt{CAFE}-fit PAH spectrum from this region as a black line, and the PAH 7.7\,\micron\ feature as a blue line. The dashed orange lines indicate the half-power of the F770W filter. Finally, the shaded orange and green regions correspond to the fractional PAH power used to compute $c_{\mathrm{PAH}}$ and $c_{\mathrm{wing}}$, respectively (see \S~\ref{sec:estimatingPAHfluxes}).}
    \label{fig:specplot}
\end{figure*}

\section{Observations and Data Reduction}\label{sec:data}

This work uses MIRI IFU spectroscopy of four LIRGs in the GOALS ERS sample: NGC\,3256, NGC\,7469, VV\,114, and II\,Zw\,96 (Science program 1328, Co-PIs: L. Armus and A. Evans). The MIRI cubes are reduced using pipeline version \Pipeline\ and with CRDS number \CRDS. After the pipeline reduction, we matched the angular resolution of the wavelength-planes of each cube to the resolution at $\lambda = 11.3\,\micron$, which is the central wavelength of the longest-wavelength PAH feature that we consider. We use the fit from \cite{law_2023}, which gives a full-width at half-maximum (FWHM) for the point spread function (PSF) of the MIRI IFU at 11.3\,\micron\ of $\sim 0\arcsec.48$ FWHM, and convolve these planes with matching Gaussian kernels of the relevant sizes.

The 4.8--28\,\micron\ MIRI/MRS spectra are extracted from a 13x13 grid of square apertures for each galaxy with a side length of $0\arcsec.4$ using the CAFE Region Extraction Tool Automaton (\texttt{CRETA})\footnote{\label{CAFEgithub}CRETA, CAFE: https://github.com/GOALS-survey/CAFE} (Diaz-Santos et al., in prep.); an example spectrum extracted from VV\,114 is shown in Figure~\ref{fig:specplot}. This extraction grid completely covers the MIRI channel 1 FOV which is the smallest of the four channels. The side length of these extraction apertures are approximately the FWHM of the PSF for the MIRI IFU at $\sim 9\,\micron$ \citep{law_2023}, which is approximately halfway between the PAH features of interest in this work (7.7 and $\mathrm{11.3\,\mu m}$). This aperture size maximizes the number of extracted spectra while still mitigating aperture effects introduced by smaller extraction regions. For each target, the center aperture of the grid is itself centered on the brightest source in the field of view, typically at the nucleus, but the two pointings of NGC 3256 (N and S) are treated separately in this regard, and we ensure there are no overlapping regions in the two pointings. To ensure good data quality, we exclude spectra containing NaNs or negative flux values.

Adopting $D=\DistNGCthree,\ \DistNGCseven,\ \DistVV$, and $\DistZw\,\mathrm{Mpc}$ \citep{armus_2009}, these $0\arcsec.4$ apertures subtend physical scales of \ScaleNGCthree, \ScaleNGCseven, \ScaleVV, and \ScaleZw\,pc for NGC\,3256, NGC\,7469, VV\,114, and II\,Zw\,96, respectively. All extracted spectra are shifted to the rest frame using the adopted redshifts $z=\RedshNGCthree,\ \RedshNGCseven,\ \RedshVV$, and $\RedshZw$, and all of the following analysis is conducted on these rest-frame spectra. Spectral decomposition is performed using the Continuum And Feature Extraction tool (\texttt{CAFE}\textsuperscript{\ref{CAFEgithub}}) \citep[Diaz-Santos et al., in prep.][]{marshall_2007}, which provides the quantities that we will compare the photometrically-inferred results to---the observed integrated flux of PAH features and the total continuum. We compare with the observed PAH flux from \texttt{CAFE} rather than the attenuation-corrected flux.

The focus of this work is on spectra from regions dominated by star-formation, so we apply cuts to exclude spectra that have a small 6.2\,\micron\ PAH equivalent width (EW), which is an indicator for a significant AGN contribution. The 6.2\,\micron\ PAH EW for each spectrum is determined using a spline continuum following the method described in \cite{spoon_2007} and spectra with $\mathrm{EW}<0.2\,\micron$ are removed. This cut is selected to primarily exclude the regions containing or adjacent to known AGN: the nucleus of NGC\,7469 \citep{landt_2008} and the southwest nucleus of VV\,114 \citep{evans_2022}. However, this cut also excludes six PAH-faint regions from II\,Zw\,96 and one from NGC\,3256\,S. These cuts result in \SpecNum\ total spectra used in the analyses described in the following sections of this paper: \SpecnumNGCthreeN, \SpecnumNGCthreeS, \SpecnumNGCseven, \SpecnumVV, and \SpecnumZw\ each from NGC\,3256\,N, NGC\,3256\,S, NGC\,7469, VV\,114, and II\,Zw\,96, respectively. 

The final sample of extracted regions spans a wide range of physical characteristics. Using a star-formation rate (SFR) prescription based on the sum of fluxes from the [NeII] (12.8\,\micron) and [NeIII] (15.5\,\micron) from \cite{whitcomb_2020}, we find that the SFR density of our regions span \SFRmin\ to \SFRMax\,$\mathrm{M_{\odot}\,yr^{-1}\,kpc^{-2}}$ with a median value of \SFRMed\,$\mathrm{M_{\odot}\,yr^{-1}\,kpc^{-2}}$. Following the method suggested in \cite{spoon_2007}, we also determine the strength of the 9.7\,\micron\ silicate absorption $S_{\mathrm{sil}}$ feature (see Figure~\ref{fig:specplot} for a moderate example) for each region, where $S_{\mathrm{sil}}=0$ indicates no absorption and the most heavily absorbed galaxies have $S_{\mathrm{sil}}\approx-4$ \citep{spoon_2007}. Our sample spans a significant portion of this parameter space, with a maximum $S_{\mathrm{sil}}=\SsilMax$, a minimum of \SsilMin, and a median value of \SsilMed.

Synthetic photometry of MIRI imaging bands is derived from each of our spectra \citep[see][]{koornneef_1986, gordon_2022}. The synthetic photometry integrations are performed only over positive spectral values using the \texttt{scipy} method \texttt{trapezoid}, and the filter transmission curves \citep{glasse_2015} are interpolated linearly between each wavelength element to match the wavelengths associated with the MIRI/MRS spectra using the \texttt{scipy} function \texttt{interp1d}. Using the synthetic photometry from MIRI filters for this analysis affords the advantage of directly comparing spectral flux density values to photometric values in a given aperture, without the complexities of the different PSFs and sensitivities between the IFU and imaging modules. As the photometry from imaging data and the synthetic photometry from spectra have principally the same value for the same aperture and filter, we hereafter refer to synthetic photometry as photometry.

\section{Analysis}

An accurate estimation of the PAH flux within a photometric band hinges upon how well the contribution from the continuum can be estimated and removed. In the mid-infrared, this continuum arises primarily from the thermal emission from large dust grains emitting at a range of temperatures, roughly forming a power-law spectral shape. Our prescription capitalizes on this to infer a pseudo-continuum at two MIRI bands centered on PAH features (F770W and F1130W, see Figure~\ref{fig:specplot}) by anchoring a power law between a choice of two continuum-dominated bands (F560W, F1000W, F1500W, and F2100W), similar to \cite{marble_2010}. We exclude F2550W because a significant portion of this bandpass is outside of the range of MIRI spectra. Here, we describe how we obtain a pseudo-continuum as well as a few constants required to extract PAH flux from MIRI photometry. In the following section, we compare the efficacy of using various combinations of MIRI bands to recover this flux for each considered PAH feature.

\subsection{Power Law Continuum}\label{sec:powerlawcont}

If $f_{\mathrm{blue}}$ and $f_{\mathrm{red}}$ are the photometric flux density measurements in two continuum-dominated bands, then the flux density associated with filter $b$ from a simple power law intersecting these measurements is 

\begin{equation} \label{eq:fcont_nocorr}
f_{\mathrm{Cont},b}\,(\mathrm{Jy}) = f_{\mathrm{blue}}^{(1-\alpha)} f_{\mathrm{red}}^{(\alpha)}\,(\mathrm{Jy}) ,
\end{equation}

where

\begin{equation} \label{eq:alpha}
\alpha = \frac{\mathrm{log}(\lambda_{b}) - \mathrm{log}(\lambda_{\mathrm{blue}})}{\mathrm{log}(\lambda_{\mathrm{red}}) - \mathrm{log}(\lambda_{\mathrm{blue}})}\ ,
\end{equation}

\noindent and $\lambda_{b}$, $\lambda_{\mathrm{red}}$, and $\lambda_{\mathrm{blue}}$ are the pivot wavelengths \citep{gordon_2022} of each filter. Thus, Eq.~\ref{eq:fcont_nocorr} gives the continuum emission within $b$. An example of such a power-law continuum evaluated over all wavelengths in the MIRI/MRS range is shown in Figure~\ref{fig:specplot}.

\subsection{Estimating PAH Fluxes}\label{sec:estimatingPAHfluxes}

The PAH 7.7\,\micron\ and PAH 11.3\,\micron\ features are usually treated as complexes consisting of multiple Drude profiles centered at 7.42, 7.60, and 7.85\,\micron\ and 11.23 and 11.33\,\micron, respectively \citep{smith_07}. Ratios of these complexes are often used to infer the properties of PAH grains in galaxies in both observations \citep[e.g.,][]{odowd_09, lai_2022} and models \citep[e.g.,][]{draine_2021}. However, other PAH features contributing to these MIRI bands need to be removed to enable a direct comparison with models and previous works: the 6.7, 8.3 and 8.6\,\micron\ features in F770W, and the 11.0\,\micron\ feature and the blue wing of the 12.0\,\micron\ feature in F1130W. 

To isolate these complexes of interest, we need to determine their fractional contribution to the total PAH emission within these bands, $c_{\mathrm{PAH}}$. The \texttt{CAFE} fit to each spectrum provides the parameters of each individual PAH feature as well as the combined PAH spectrum \citep[see][]{lai_2022, bohn_2024}. For each spectrum, we take the ratio of the integrated power of a PAH complex of interest to that of the total PAH spectrum within a given filter bandpass based on the fits. See the inset plot in Figure~\ref{fig:specplot} for an example of this fractional PAH power. Taking the median of these over all spectra gives $c_{\mathrm{PAH}}$ for a given complex of interest, which are given for each complex in Table~\ref{table:constants}. Although the relative strengths of PAH features vary according to environmental factors as well as characteristics of the distribution of PAH size and ionization, the magnitude of this variation is typically insignificant for these adjacent features. The variation in $c_{\mathrm{PAH}}$ across our sample amounts to $\lesssim 2 \%$ for the 7.7\,\micron\ and 11.3\,\micron\ PAH complexes.

Drude profiles have a significant fractional power contained in their broad wings. As a consequence, a non-negligible amount of the power from a rest-frame PAH complex lies outside of the corresponding MIRI filter, which we correct for with $c_{\mathrm{wing}}$. We correct for this with a similar method as was used to determine $c_{\mathrm{PAH}}$, except that the average fraction that we are interested in here is that of the total emission of that complex across all wavelengths to the PAH complex emission contained within the filter (see inset plot in Figure~\ref{fig:specplot}). This is averaged for the \texttt{CAFE} fits over all of our spectra to find $c_{\mathrm{wing}}$ for each PAH complex, which varies by 1\% across all spectra. The values of $c_{\mathrm{wing}}$ are given in Table~\ref{table:constants}. Thus, $c_{\mathrm{wing}}$ represents the boost needed to account for the fraction of PAH power lost outside of the bandpass.

The final constant accounts for the bandwidth and shape of the PAH-dominated filter $b^{\prime}$ when converting from the flux density units of the synthetic photometry to the inferred integrated flux. We determine this value $w_{b^{\prime}}$ empirically as the median over all spectra of the ratio

\begin{equation} \label{eq:wb}
w_{b^{\prime}} = \mathrm{Med} \left( \frac{F_{\mathrm{PAH, Spec}}\,(\mathrm{W\,m^{-2}})}{10^{-26} \nu_{b^{\prime}} f_{\mathrm{PAH},b^{\prime}} \, (\mathrm{Jy})} \right)\, ,
\end{equation}

\noindent where $F_{\mathrm{PAH, Spec}}$ is the spectrally integrated flux of a PAH complex taken from \texttt{CAFE}, $\nu_{b^{\prime}}$ is the frequency associated with the pivot wavelength of $b^{\prime}$, and $f_{\mathrm{PAH},b^{\prime}}$ is the synthetic photometry for the model of this PAH complex from \texttt{CAFE} within band $b^{\prime}$. The values of $w_{b^{\prime}}$ are given in Table~\ref{table:constants}.

\begin{table}[h!]
\centering
\begin{tabular}{@{} l c c c c @{}}
\toprule
\textbf{Complex} & \textbf{$b^{\prime}$} & \textbf{$w_{b^{\prime}}$} & \textbf{$c_{\mathrm{PAH}}$} & \textbf{$c_{\mathrm{wing}}$} \\
\midrule
PAH $\mathrm{7.7\,\mu m}$ & F770W & $0.271$ & $0.74 \pm 0.01$ & $1.15 \pm 0.01$ \\ 
PAH $\mathrm{11.3\,\mu m}$ & F1130W & $0.066$ & $0.87 \pm 0.02$ & $1.16 \pm 0.01 $ \\ 
\bottomrule
\end{tabular}
\caption{Summary of derived constants and their values for the 7.7\,\micron\ and 11.3\,\micron\ PAH complexes. For a PAH-dominated MIRI filter $b^{\prime}$, $w_{b^{\prime}}$ accounts for the filter width and shape, $c_{\mathrm{PAH}}$ isolates the flux from the PAH complex of interest from the total PAH flux in $b^{\prime}$ , and $c_{\mathrm{wing}}$ corrects for lost flux of the PAH complex of interest out of $b^{\prime}$ (see \S~\ref{sec:estimatingPAHfluxes} for each). Reported uncertainties are $1\sigma$ from all spectra in our sample.}
\label{table:constants}
\end{table}

Combining the previously described components, the flux of a given PAH complex is estimated from MIRI photometry as

\begin{equation} \label{eq:fpah_uncal}
\begin{aligned}
F_{\mathrm{PAH},b^{\prime}}\, (\mathrm{W m^{-2}}) = 10^{-26}\, \nu_{b^{\prime}}\, w_{b^{\prime}}\, c_{\mathrm{wing}}\, c_{\mathrm{PAH}}\,\times \\
    (f_{b^{\prime}} - f_{\mathrm{blue}}^{(1-\alpha)} f_{\mathrm{red}}^{(\alpha)})
\end{aligned}
\end{equation}

\noindent for flux densities $f$ given in Jy.

\section{Results}

In this section, we apply our prescription via Eq.~\ref{eq:fpah_uncal} to several combinations of JWST/MIRI filters to extract the PAH 7.7\,\micron\ and PAH 11.3\,\micron\ fluxes, and we compare the resulting flux values to those from the \texttt{CAFE} fits. Then, we introduce a calibration to the inferred continuum derived from comparisons to spectral fits, and compare the resulting new PAH flux estimates to values from the \texttt{CAFE} fits. In general, we evaluate the performance of a given set of continuum bands based upon two criteria: the systematic offset of the inferred PAH flux relative to the spectroscopically measured \texttt{CAFE} PAH flux, and the spread of the distribution around this offset. These are quantified by the median of the distribution of percent differences with the spectroscopic flux, and the 25--$75^{\mathrm{th}}$ percentile range (the inter-quartile range or IQR) of this distribution, respectively. 

\subsection{Uncorrected Photometric PAH Estimates}

\begin{figure*}[htb]
    \includegraphics[width=1\textwidth]{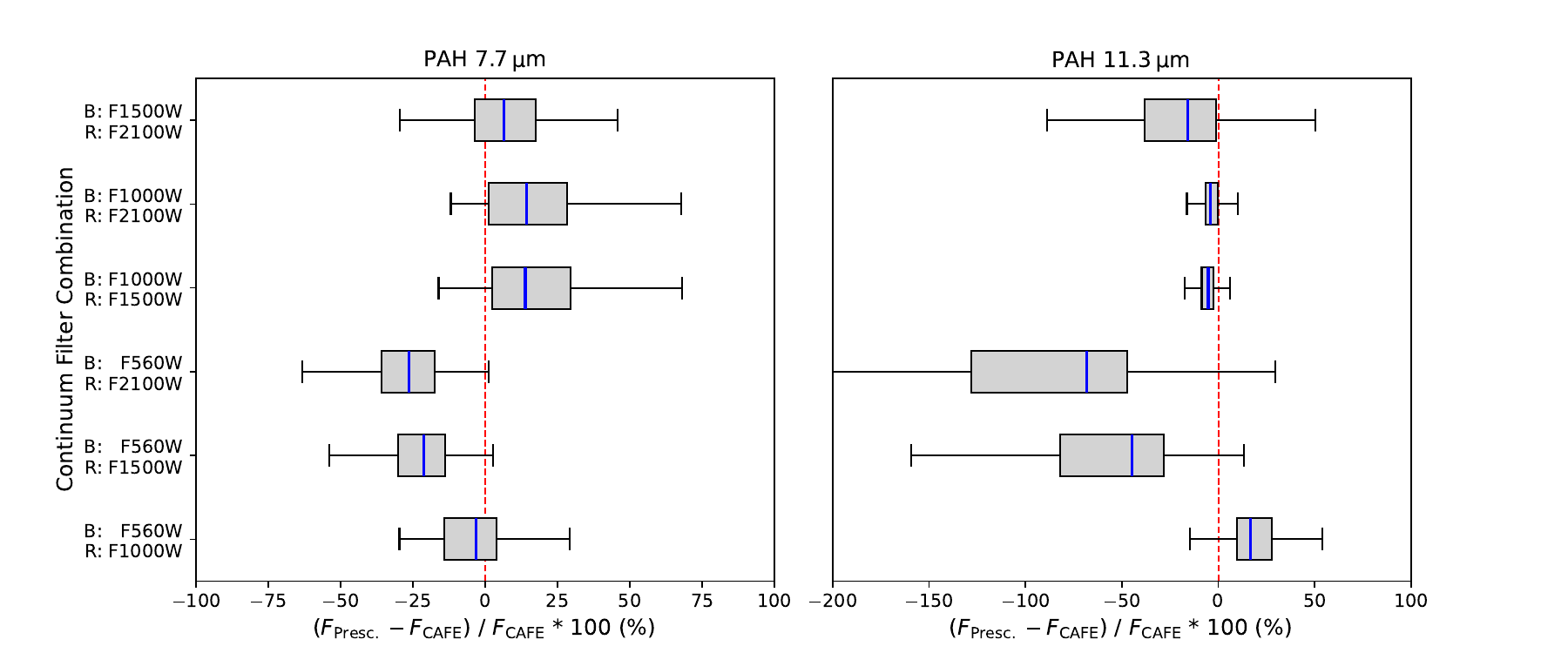}
    \caption{Comparisons between inferred photometric PAH flux and PAH feature strength measured from the JWST spectra with \texttt{CAFE} for various combinations of JWST/MIRI filters using Eq.~\ref{eq:fpah_uncal}. Characteristics of the distribution expressed as a percent difference from the \texttt{CAFE} values are shown, with the medians as blue lines, the $25^{\mathrm{th}}$ and $75^{\mathrm{th}}$ percentiles as the left and right of each gray box, respectively, and the horizontal black lines indicate the largest percent difference within 1.5 times the 25--$75^{\mathrm{th}}$ percentile range on either side. The horizontal range in the right panel is fixed with -200\% on the left for clarity. The left panel corresponds to values for PAH 7.7\,\micron, and the right panel is for PAH 11.3\,\micron. There are both random and systematic offsets for the distributions of all filter combinations that can be greatly reduced through the use of spectroscopically derived calibrations.}
    \label{fig:no_g_barwhisker}
\end{figure*}

\begin{figure*}[htb!]
    \includegraphics[width=1\textwidth]{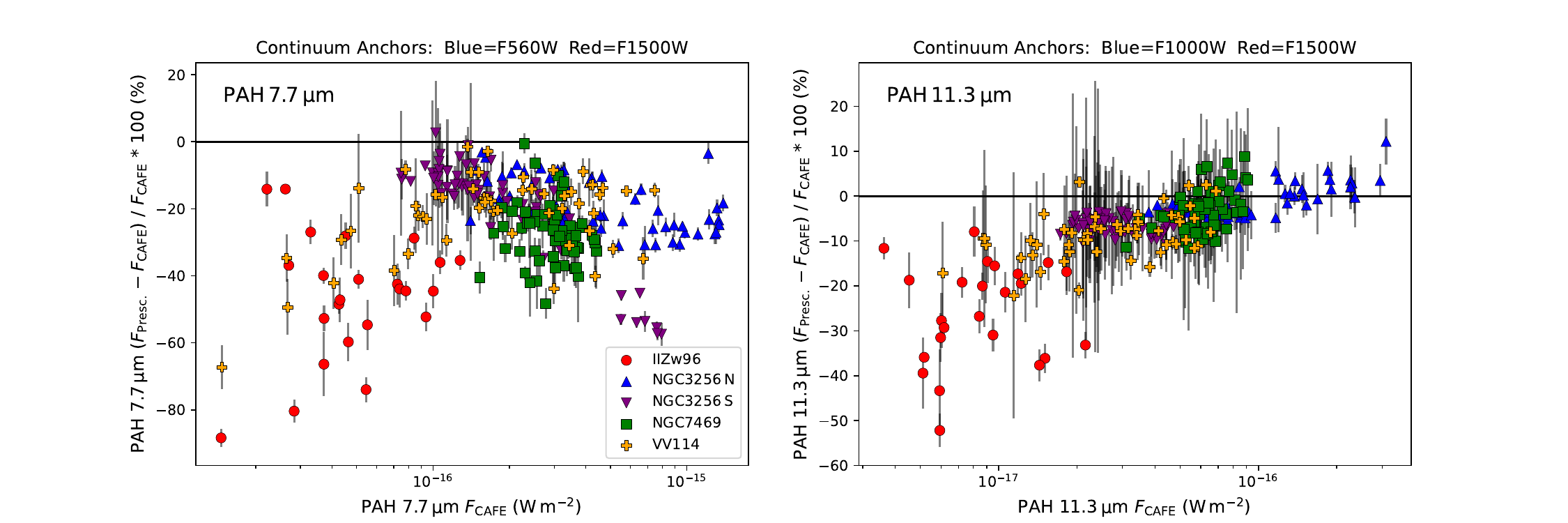}
    \caption{The percentage differences between PAH flux as estimated by Eq.~\ref{eq:fpah_uncal} and as fit with \texttt{CAFE} for PAH 7.7\,\micron\ (left) and PAH 11.3\,\micron\ (right). Colored points indicate individual spectra in our sample. The horizontal black line indicates a perfect match between photometrically and spectroscopically derived PAH fluxes. These filter combinations, indicated at the top of each panel, yield the most accurate results when a spectroscopic calibration is applied (see \S~\ref{sec:correctingEstimates}). However, they produce large differences with the \texttt{CAFE} values here without a calibration.}
    \label{fig:no_g_distros}
\end{figure*}

In Figure~\ref{fig:no_g_barwhisker}, we show the results from Eq.~\ref{eq:fpah_uncal} for each set of continuum bands and both PAH complexes of interest compared with the \texttt{CAFE} values. For the PAH 7.7\,\micron\ complex shown in the left panel of Figure~\ref{fig:no_g_barwhisker}, the best-performing band combination is Blue:F560W, Red:F1000W. For the PAH 11.3\,\micron\ complex, there are two combinations that produce similarly accurate results: Blue:F1000W, Red:F1500W, and Blue:F1000W, Red:F2100W, with the latter yielding a slightly smaller systematic offset. 

The right panel of Fig.~\ref{fig:no_g_barwhisker} shows that the F1000W band is critical for an accurate estimation of the PAH 11.3\,\micron\ flux. This results from the PAH 11.3\,\micron\ complex being situated within the absorption feature centered at 9.7\,\micron\ and attributed to silicate grains (\cite{kemper_2004}, see Figure~\ref{fig:specplot}); the F1000W band is necessary to trace the depth of this absorption feature and establish the blue end of the continuum. 

The percent difference of the flux inferred from photometry and of that from the \texttt{CAFE} fit for each spectrum is shown in Figure~\ref{fig:no_g_distros} for certain filter combinations, Blue:F560W, Red:F1500W for PAH 7.7\,\micron\ and Blue:F1000W, Red:F1500W for PAH 11.3\,\micron. These combinations perform the best after a calibration is applied (see \S\ref{sec:correctingEstimates}\ \& \S\ref{sec:discussion}), so we show the uncalibrated results in Figure~\ref{fig:no_g_distros} for later comparison. The photometric estimate of the PAH 7.7\,\micron\ complex using this combination is underestimated for almost all regions, typically by about $21\%$. For the PAH 11.3\,\micron\ complex in the right panel, the uncalibrated estimates for this combination are accurate to within about 5\% for the brightest PAH fluxes ($F_{\mathrm{CAFE}}>5\times 10^{-17}\ \mathrm{W m^{-2}}$), but the PAH flux is increasingly underestimated up to 25--50\% for regions with fainter PAH 11.3\,\micron\ flux.

\subsection{Correcting Photometric PAH Estimates}\label{sec:correctingEstimates}

\begin{table*}[htb]
\centering
\begin{tabular}{c@{\hskip 0.15cm}c@{\hskip 0.15cm}c@{\hskip 0.15cm}c@{\hskip 0.5cm}c@{\hskip 0.15cm}c@{\hskip 0.15cm}c@{\hskip 0.15cm}c@{\hskip 0.15cm}c}
& \multicolumn{4}{c}{\textbf{$\mathrm{PAH\ 7.7\mu m\ Complex}$}} & \multicolumn{4}{c}{\textbf{$\mathrm{PAH\ 11.3\mu m\ Complex}$}} \\
\cmidrule(r){2-5} \cmidrule(l){6-9}
\textbf{Filters (B, R)} & \textbf{$\alpha$} & \textbf{$g_{\mathrm{cont}}$} & \textbf{No Cal. (\%)} & \textbf{Cal. (\%)} & \textbf{$\alpha$} & \textbf{$g_{\mathrm{cont}}$} & \textbf{No Cal. (\%)} & \textbf{Cal. (\%)} \\
\midrule
F1500W, F2100W & -2.11 & $1.29$ & $6.6\,(-3.7, 17.5)$ & $-1.5\,(-17.2, 10.6)$ & -0.89 & $0.81$ & $-15.7\,(-38.2, -1.1)$ & $4.3\,(-13.3, 13.8)$ \\ 
F1000W, F2100W & -0.36 & $1.51$ & $14.3\,(1.0, 28.4)$ & $2.7\,(-14.5, 18.3)$ & 0.17 & $1^{\dag}$ & $-4.0\,(-6.8, -0.1)$ & $-4.0\,(-6.8, -0.1)$ \\ 
F1000W, F1500W & -0.64 & $1.56$ & $13.8\,(2.4, 29.5)$ & $1.7\,(-16.4, 19.6)$ & 0.31 & $1^{\dag}$ & $-5.2\,(-8.6, -2.5)$ & $-5.2\,(-8.6, -2.5)$ \\ 
F0560W, F2100W & 0.23 & $0.62$ & $-26.4\,(-35.8, -17.5)$ & $-1.1\,(-7.7, 6.4)$ & 0.53 & $0.49$ & $-68.0\,(-128.0, -47.2)$ & $5.2\,(-21.8, 19.2)$ \\ 
F0560W, F1500W & 0.31 & $0.68$ & $-21.3\,(-30.0, -13.8)$ & $-0.6\,(-7.7, 6.2)$ & 0.71 & $0.6$ & $-44.6\,(-81.8, -28.3)$ & $5.4\,(-19.4, 16.0)$ \\ 
F0560W, F1000W & 0.53 & $0.91$ & $-3.2\,(-14.1, 3.9)$ & $0.4\,(-9.7, 8.0)$ & 1.22 & $1.18$ & $16.7\,(9.5, 27.5)$ & $4.1\,(-2.6, 14.0)$ \\
\bottomrule
\end{tabular}
\caption{The $\alpha$ for each filter combination is computed via Eq.~\ref{eq:alpha}, and the continuum correction $g_{\mathrm{cont}}$ is reported for each filter combination as the median ratio of the continuum from \texttt{CAFE} to the photometric psuedo-continuum computed from all spectra in our sample. The columns labeled No Cal. and Cal. correspond to the median percent difference between the PAH flux inferred from photometry vs. fit with \texttt{CAFE} (as photometric~-~\texttt{CAFE}) without and with $g_{\mathrm{cont}}$ applied, respectively. These are given with the 1st (left) and 3rd (right) quartile percent differences. Entries marked with \dag\ indicate that having no continuum correction ($g_{\mathrm{cont}}=1$) results in more accurate fluxes for PAH\,11.3\,\micron. See \S~\ref{sec:discussion} for an alternative prescription for these filter combinations.}
\label{table:performance}
\end{table*}

In our sample of spectra, there is an overall offset between the PAH fluxes measured with \texttt{CAFE} and the fluxes estimated by Eq.~\ref{eq:fpah_uncal} for a given set of continuum bands, see Figure~\ref{fig:no_g_barwhisker}. One approach to account for this this would be to simply apply a correction for each filter combination directly to the resulting PAH flux in order to align the median of each distribution to 0\% difference with \texttt{CAFE} values, but that would preserve the same (often large) IQR within each distribution. Instead, we introduce a correction factor $g_{\mathrm{cont}}$ that can be applied to the photometric power-law continuum within each region. This is done by finding the median relative offset between the value of the inferred continuum at the fiducial wavelength of a PAH-dominated band, and the synthetic photometry of the featureless \texttt{CAFE} fitted continuum in that band. For example, this would correspond to the ratio between the solid red line and the dashed red line at 7.7\,\micron\ for PAH 7.7\,\micron, Blue:F560W, Red:F1500W in Figure~\ref{fig:specplot}. Table~\ref{table:performance} gives $g_{\mathrm{cont}}$ for each PAH complex and filter combination. Using these, the spectroscopically-calibrated versions of our prescription become

\begin{equation} \label{eq:fpah77_cal}
\begin{aligned}
F_{\mathrm{PAH\,7.7}}\, (\mathrm{W m^{-2}}) = (9.00\pm0.21)\times10^{-14}\,\mathrm{Hz}\ \times\,\\
(f_{b^{\prime}} - g_{\mathrm{cont}}\, f_{\mathrm{blue}}^{(1-\alpha)} f_{\mathrm{red}}^{(\alpha)})
\end{aligned}
\end{equation}

and

\begin{equation} \label{eq:fpah11_cal}
\begin{aligned}
F_{\mathrm{PAH\,11.3}}\, (\mathrm{W m^{-2}}) = (1.77\pm0.04)\times10^{-14}\,\mathrm{Hz}\ \times\,\\
(f_{b^{\prime}} - g_{\mathrm{cont}}\, f_{\mathrm{blue}}^{(1-\alpha)} f_{\mathrm{red}}^{(\alpha)})
\end{aligned}
\end{equation}

\noindent for values of flux densities $f$ given in Jy. Find values for $\alpha$ and $g_{\mathrm{cont}}$ in Table~\ref{table:performance}. Note that Eq.~\ref{eq:fpah11_cal} does not account for the 9.7\,\micron\ silicate absorption feature, and an additional correction is necessary for significantly absorbed sources (see \S~\ref{sec:discussion}).

\begin{figure*}[htb]
    \includegraphics[width=1\textwidth]{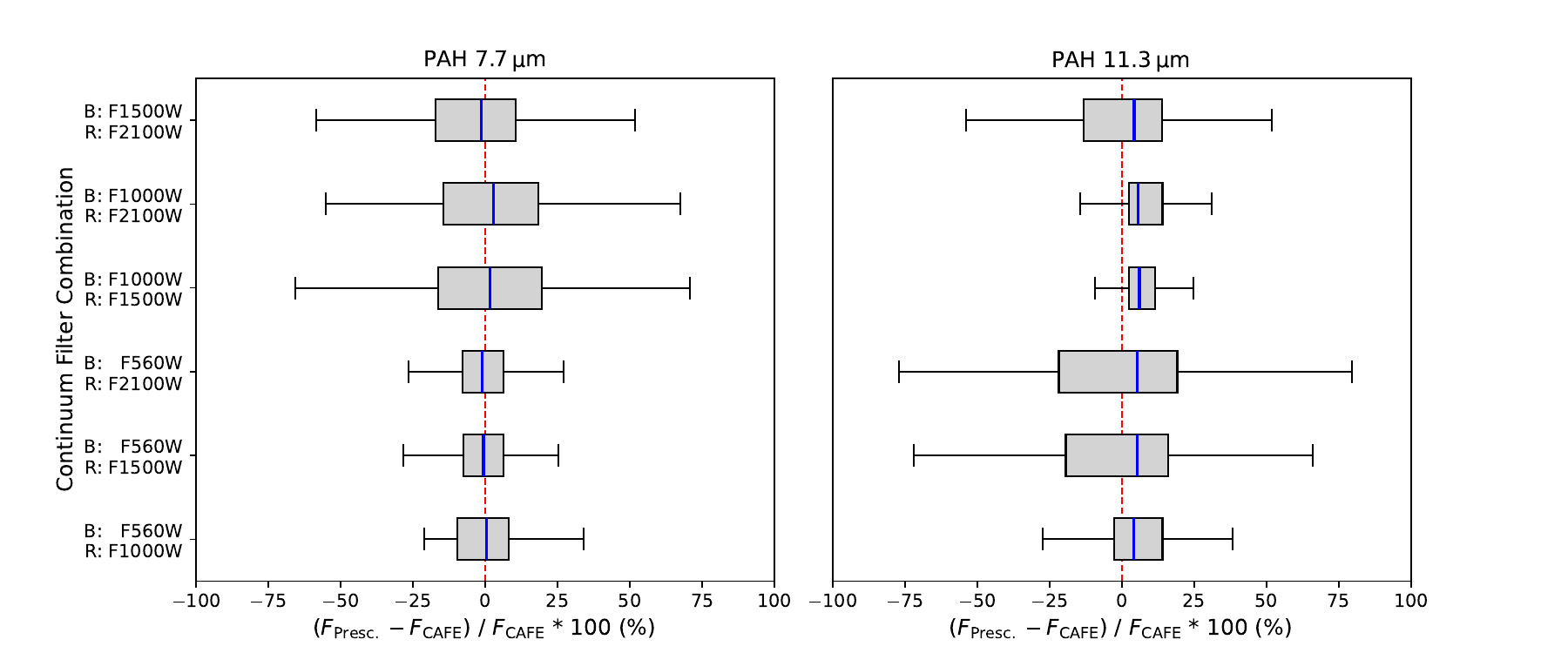}
    \caption{The same as Fig.~\ref{fig:no_g_barwhisker}, except the photometric PAH fluxes were estimated using Eq.~\ref{eq:fpah77_cal} (left) and Eq.~{\ref{eq:fpah11_cal}} (right), meaning the continuum correction $g_{\mathrm{cont}}$ was applied. For B:F1000W, R:F1500W and B:F1000W, R:F2100W with PAH 11.3\,\micron, we use $g_\mathrm{cont}=0.86$ and 0.88, respectively, as computed via the method described in \S~\ref{sec:correctingEstimates}. Comparing with Fig.~\ref{fig:no_g_barwhisker}, the overall offsets have been largely removed and the IQR has been reduced for most filter combinations.}
    \label{fig:with_g_barwhisker}
\end{figure*}

\begin{figure*}
    \includegraphics[width=1\textwidth]{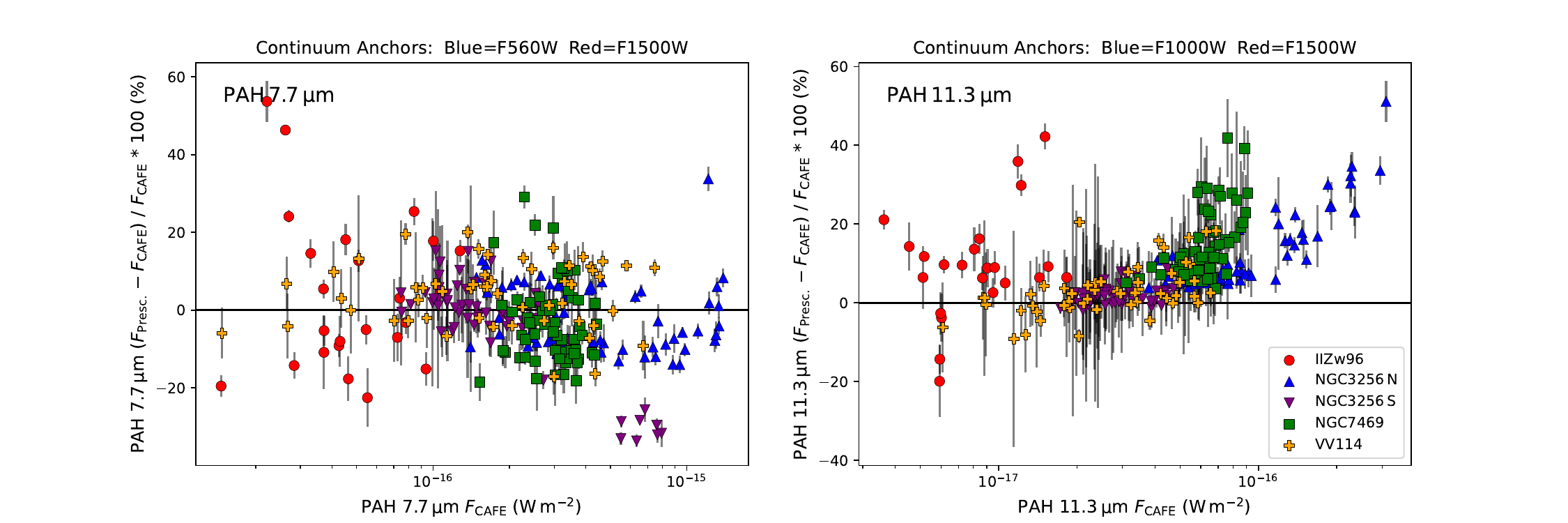}
    \caption{The same as Fig.~\ref{fig:no_g_distros}, except the photometric PAH fluxes were estimated including the $g_{\mathrm{cont}}$ correction using Eq.~\ref{eq:fpah77_cal} (left) and Eq.~{\ref{eq:fpah11_cal}} (right). In the right panel, $g_{\mathrm{cont}}=0.86$ was used, instead of $g_{\mathrm{cont}}=1$ as in Table~\ref{table:performance} (see text). There is an overall improvement for PAH 7.7\,\micron\ compared with Fig.~\ref{fig:no_g_distros}. For PAH 11.3\,\micron, the apparent trend of underestimated prescribed PAH fluxes is removed, but there is an overestimated flux for bright spectra with a minimal silicate absorption feature. Filter combinations using Blue:F1000W for PAH 11.3\,\micron\ require an adjusted prescription (see \S~\ref{sec:discussion}).} 
    \label{fig:with_g_distros}
\end{figure*}

In Fig.~\ref{fig:with_g_barwhisker}, we once again show the performance of each set of continuum bands compared to the spectral fits from \texttt{CAFE} for extracting PAH fluxes, but with the correction factor for each combination applied. The systematic offsets with \texttt{CAFE} values for both PAH complexes are greatly reduced, and are essentially eliminated for all filter combinations in the case of PAH 7.7\,\micron. There is a remaining systematic offset that is shared between all of the filter combinations for the estimation of the PAH 11.3\,\micron\ complex, and critically, the two previously best-performing combinations (Blue:F1000W, and F1500W or F2100W) perform worse after the calibration in both median offset and IQR. Thus, we artifically set $g_{\mathrm{cont}}=1$ for these combinbations in Table~\ref{table:performance}, and these combinations are further discusssed in \S~\ref{sec:discussion}. However, the IQR of the distribution for most of the filter combinations has also been reduced for both PAH complexes. This reduction is most striking for the estimation of PAH 11.3\,\micron\ by Blue:F560W, Red:F1500W and Blue:F560W, Red:F2100W. Without the continuum correction, these combinations resulted in PAH flux estimations with unacceptably large errors with median percent differences of -45 and -68\%, respectively. With the correction, they can be used with typical differences from the spectroscopically derived true fluxes of $\lesssim20\%$.

The post-correction distributions for each combination show that there is a filter for each PAH complex that will substantially improve its estimation. For the PAH 7.7\,\micron\  complex, all of the combinations that include F560W outperform all of the combinations which do not by exhibiting a smaller IQR, showing how important F560W is for establishing the blue-end of the continuum that lies within the F770W filter. Similarly, the F1000W filter is critical for the best estimation of the PAH 11.3\,\micron\ complex for essentially the same reason: to establish the blue end of the continuum that lies within the F1130W band. The complex spectral shape introduced by the silicate absorption feature at 9.7\,\micron\ magnifies the importance of the F1000W band in highly obscured regions. 

Fig.~\ref{fig:with_g_distros} is similar to Fig.~\ref{fig:no_g_distros}, but now shows the results applying the spectroscopically based correction $g_{\mathrm{cont}}$. For PAH 7.7\,\micron, the distribution of percent difference is now centered around 0\%, whereas it was previously centered around -21\%. Additionally, the IQR has been reduced by more than a factor of 3. It is worth noting, however, that Blue:F560W, Red:1000W was only slightly improved by $g_{\mathrm{cont}}$, and it performs only slightly worse than Blue:F560W, Red:F1500W in both median offset and IQR (see Table~\ref{table:performance}).

In the right panel of Fig.~\ref{fig:with_g_distros} for PAH 11.3\,\micron, we show the results of Eq.~\ref{eq:fpah11_cal} where $g_{\mathrm{cont}}=0.86$ as computed in the way described in \S~\ref{sec:correctingEstimates}, instead of $g_{\mathrm{cont}}=1$ as shown in Table~\ref{table:performance}. Here, $g_{\mathrm{cont}}$ benefited regions with fainter PAH flux by removing the trend that was apparent in the right panel of Figure~\ref{fig:no_g_distros}. However, Fig.~\ref{fig:with_g_distros} shows a new trend of growing disagreement as the \texttt{CAFE} PAH flux values increase. This offset can be removed by applying an additional correction based on the silicate optical depth, as we explain in \S\ref{sec:discussion}.

\section{Discussion}\label{sec:discussion}

PAH ratios can be used to probe the average size and ionization distributions of PAHs \citep{draine_2007, draine_2021}, and PAH emission extracted from MIRI images can be used to map these parameters over large solid angles and at high angular resolution. Using band ratios for these purposes would be particularly effective when combined with the highly grain size-sensitive PAH 3.3\,\micron\ feature extracted from NIRCam photometry. Through the spatially-resolved diversity of environments offered by the GOALS ERS sample of LIRGs, we have shown that this method is applicable for regions across a wide variety of spectral characteristics such as star-formation rate and attenuation. 

\begin{figure*}[htb!]
    \includegraphics[width=1\textwidth]{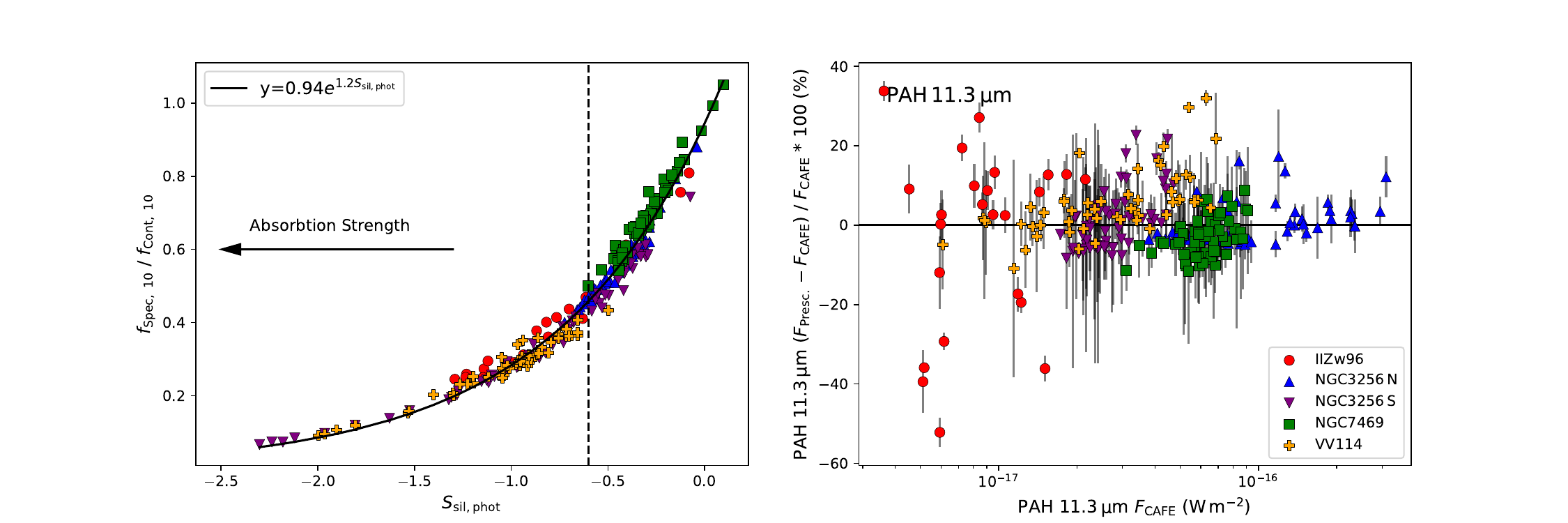}
    \caption{Left: the relationship between the photometrically-inferred silicate absorption strength $S_{\mathrm{sil, phot}}$ and the spectral flux density at 10\,\micron\ ($f_{\mathrm{Spec, 10}}$) relative to the 10\,\micron\ power law continuum anchored at F560W and F1500W (ex. red dashed line in Figure~\ref{fig:specplot}) for each region (filled shapes). This relationship can be applied to infer $f_{\mathrm{Spec, 10}}$ from F1000W photometry, which is useful for estimating the continuum in absorbed sources. To the left of the black dashed line ($S_{\mathrm{sil, phot}}=-0.6$), we correct for the effect of silicate absorption (Eq.~\ref{eq:fpah_10umfix}) for the results in the right panel. To the right of this line, no correction is applied (Eq.~\ref{eq:fpah11_cal}). Right: the same as the right panels of Figures~\ref{fig:no_g_distros} or \ref{fig:with_g_distros}, but using either Eq.~\ref{eq:fpah11_cal} or Eq.~\ref{eq:fpah_10umfix} to estimate PAH 11.3\,\micron\ flux, depending on the degree of silicate strength. The systematically underestimated fluxes from Figure~\ref{fig:no_g_distros} have been corrected, and this adjusted prescription also avoids the overestimated PAH fluxes present from $g_{\mathrm{cont}}$ in Figure.~\ref{fig:with_g_distros}.}
    \label{fig:distro_10umfix}
\end{figure*}

The flux of the 7.7\,\micron\ and 11.3\,\micron\ PAH complexes can be recovered from MIRI photometry accurate to $\sim7\%$ and $\sim5\%$ when the recommended filter combinations are used with Eq.~\ref{eq:fpah77_cal} and Eq.~\ref{eq:fpah11_cal}, respectively. These photometric accuracies are comparable to the uncertainties in the spectroscopic fitting, which are typically 5\%.

The efficacy of this spectroscopic correction ($g_{\mathrm{cont}}$) depends upon the choice of continuum filters. In general, the calibration improves the systematic and random offsets of the photometrically derived fluxes for most combinations, sometimes transforming combinations with previously unsatisfactory results into those providing an IQR of $\sim20\%$. 

When a correction to the continuum is applied as in the right panel of Figure~\ref{fig:with_g_distros}, the low attenuation, high PAH flux spectra that were previously quite accurately estimated with respect to PAH 11.3\,\micron\ flux (see Figure~\ref{fig:no_g_distros}) are now overestimated. This is due to the fact that $g_{\mathrm{cont}}$ represents the median continuum correction necessary for our sample of spectra that is affected by a wide range of silicate absorption strengths. For low or attenuation-free spectra such as those with the brightest PAH 11.3\,\micron\ from NGC\,7469 or NGC\,3256\,N, $g_{\mathrm{cont}}$ reduces the flux of the already well-estimated continuum, causing an overestimation of PAH flux that rises as the degree of absorption decreases. To illustrate this, if Eq.~\ref{eq:fpah_uncal} (no $g_{\mathrm{cont}}$ applied, Blue:F1000W, Red:F1500W) is used to estimate PAH 11.3\,\micron\ for regions with low to moderate $S_{\mathrm{sil}}>-0.5$, the accuracy is improved to $\PAHelevenLowSilMed\,(\PAHelevenLowSilIQRhigh, \PAHelevenLowSilIQRlow)\%$ preserving the format used in Table~\ref{table:performance}.

When deriving the 11.3\,\micron\ PAH flux, silicate absorption can significantly affect the results for some continuum filter combinations (Blue:F1000W, Red:F1500W and Blue:F1000W, Red:F2100W). The apparent trend of underestimation seen in Figure~\ref{fig:no_g_distros} for PAH 11.3\,\micron\ results from more absorbed sources tending to have a fainter PAH flux. Indeed, NGC\,3256\,S, VV\,114, and II\,Zw\,96 contribute the bulk of regions with underestimated PAH flux in the right panel of Figure~\ref{fig:no_g_distros} and contain the most highly obscured spectra (mean $S_{\mathrm{sil}}=\AvSsilNGCthirtyS, \AvSsilVV, \AvSsilZw$, respectively), while the relatively low-absorption NGC\,7469 and NGC\,3256\,N (mean $S_{\mathrm{sil}}=\AvSsilNGCseven, \AvSsilNGCthirtyN$, respectively) do not show this underestimation. By including an extra filter in the prescription, we can increase the accuracy in the face of significant dust attenuation. Here, we provide an adjusted prescription for these filter combinations to estimate PAH 11.3\,\micron\ for sources with a significant silicate absorption strength.

The F1000W band is wide enough to encompass a significant amount of continuum on either side of the wavelength of maximum absorption (9.7\,\micron). When the absorption is strong, the synthetic photometry of F1000W yields higher flux values than the actual spectrum at 10\,\micron\, as demonstrated by the example in Figure~\ref{fig:specplot}. This leads to an overestimation of an inferred continuum anchored at F1000W, and thus an underestimation of the PAH flux. Since the shape of this absorption feature is relatively constant and it primarily varies in its strength, the offset between the synthetic photometry for F1000W, $f_{\mathrm{Phot,10}}$, and the spectral flux density at 10\,\micron, $f_{\mathrm{Spec,10}}$, both relative to the unabsorbed continuum, $f_{\mathrm{Cont,10}}$, should vary predictably with the silicate strength, $S_{\mathrm{sil}}$. We can estimate $f_{\mathrm{Cont,10}}$ from photometry by applying Eq.~\ref{eq:fcont_nocorr} with Blue:F560W, Red:F1500W, evaluated for 10\,\micron. In Figure~\ref{fig:specplot}, $f_{\mathrm{Cont,10}}$ estimated this way would appear on the red dashed line at 10\,\micron. Similarly, we estimate the silicate strength photometrically as 

\begin{equation} \label{eq:S_sil}
S_{\mathrm{sil,\,phot}} = \mathrm{ln} \left[ \frac{f_{F1000W}}{f_{F560W}^{0.42} \, f_{F1500W}^{0.58}}  \right] \, .
\end{equation}

Using this combination of anchor bands to probe the silicate absorption rather than the more nearby Blue:F770W, Red:F1130W provides a better estimate of the PAH and true continuum level between the 7.7 and 11.3\,\micron\ PAH emission features. We find that there is a close linear relation between $S_{\mathrm{sil,\,phot}}$ and $S_{\mathrm{sil}}$ from the method outlined in \cite{spoon_2007} (see \S~\ref{sec:data}), $S_{\mathrm{sil}} = 1.29S_{\mathrm{sil,\,phot}} + 0.24$, with a Pearson correlation coefficient $r = 0.99$.

The relationship between $S_{\mathrm{sil,\,phot}}$ and $f_{\mathrm{Spec,10}}/f_{\mathrm{Cont,10}}$ takes the form of a simple exponential function as shown in the left panel of Figure~\ref{fig:distro_10umfix}. With this relation, the spectral flux value at 10\,\micron\ can be estimated from the photometry of F560W, F1000W, and F1500W. Substituting this inferred 10\,\micron\ spectral flux for $f_{\mathrm{blue}}$ in Eq.~\ref{eq:fpah_uncal}, the adjusted prescription for PAH 11.3\,\micron\ with Blue:F1000W becomes

\begin{equation} \label{eq:fpah_10umfix}
\begin{aligned}
F_{\mathrm{PAH\,11.3}}\, (\mathrm{W m^{-2}}) = (1.77\pm0.04)\times10^{-14}\,\mathrm{Hz}\, \times \\
    \left[ f_{b^{\prime}} - \left\{ 0.94\,f_{\mathrm{\mathrm{F1000W}}}^{1.20} \,f_{\mathrm{F560W}}^{-0.08}\,f_{\mathrm{F1500W}}^{-0.12} \right\}^{(1-\alpha)} f_{\mathrm{red}}^{(\alpha)} \right] \, ,
\end{aligned}
\end{equation}

where $f_{\mathrm{red}}$ is the flux density in either F1500W or F2100W and all flux densities have units of Jy. Testing $S_{\mathrm{sil,\,phot}}$ cutoffs between -2.5 and 0, we find that Eq.~\ref{eq:fpah_10umfix} performs better than Eq.~\ref{eq:fpah11_cal} when there is significant absorption of $S_{\mathrm{sil,\,phot}}<-0.6$. For reference, this corresponds to $S_{\mathrm{sil}}\approx-0.54$ from the method of \cite{spoon_2007}.

When Eq.~\ref{eq:fpah11_cal} and Eq.~\ref{eq:fpah_10umfix} are used in tandem to estimate PAH 11.3\,\micron\ for sources with $S_{\mathrm{sil,\,phot}}>-0.6$ and $<-0.6$, respectively, the median and IQR percent differences with CAFE across all regions in our sample are $-0.3,(-4.2, 4.3)\%$ for Red:F1500W and $2,(-5.0, 10.5)\%$ for Red:F2100W, presented in the same format used in Table~\ref{table:performance}. For Red:F1500W, the overall offset has been removed while preserving a similar IQR as before. For Red:F2100W, the distribution of percent differences with \texttt{CAFE} is better centered around $0\%$, but the IQR has increased. Figure~\ref{fig:distro_10umfix} shows the results for Red:F1500W where Eq.~\ref{eq:fpah11_cal} and Eq.~\ref{eq:fpah_10umfix} are applied selectively based on $S_{\mathrm{sil,\,phot}}$. Here, we see the under/overestimation trends in the right panels of Figures~\ref{fig:no_g_distros} and \ref{fig:with_g_distros} have been removed. This adjusted prescription for PAH 11.3\,\micron\ accounts for the complexity introduced by the silicate absorption.

This work focuses on estimating the \textit{observed} PAH flux from MIRI photometric data. Various spectral decomposition tools/methods attempt to recover the true or \textit{intrinsic} flux by including a MIR attenuation law as a scalable or fittable component and assuming a simple geometry for the obscuring dust \citep[e.g.][]{smith_07, marshall_2007, stierwalt_2014, Xie_2018}. See \cite{lai_2024} for a comparison of various MIR attenuation laws and geometry scenarios. By determining the degree of attenuation with $S_{\mathrm{sil,\,phot}}$, it may be possible to estimate the intrinsic PAH flux photometrically in a similar manner. In highly obscured environments, such a correction would be necessary to accurately infer PAH properties.

The prescriptions for both the PAH 7.7 and 11.3\,\micron\ complexes are intended for photometric measurements of targets in the local universe. To quantify at what maximum redshift the flux errors become unacceptably large, we applied an artificial redshift to three spectra from our sample, re-computed the synthetic photometry, and compared the results of our prescription for redshifted to spectra to the results at rest. This was done to $z=0.06$, which is approximately where the PAH 11.3\,\micron\ complex is completely shifted out of F1130W. We find that the estimated flux for PAH 7.7\,\micron\ is underestimated by $\sim2\%$ at $z=0.01$, $\sim7\%$ at $z=0.03$, and between 8 and 16\% by $z=0.06$, depending on the spectrum. For low-attenuation spectra, the flux of PAH 11.3\,\micron\ stays within $\pm2.5\%$ until $z\approx0.015$, at which point it is underestimated rapidly to $\sim-30\%$ at $z=0.03$ and $\sim-80\%$ at $z=0.06$. 

\section{Summary}

In this work, we detail a prescription for estimating the flux of the PAH 7.7\,\micron\ and PAH 11.3\,\micron\ complexes from MIRI photometry for low-redshift sources over a wide range of star-formation rate, dust attenuation, and spectral slope. This prescription is calibrated using \SpecNum\ independent spectra from four star forming luminous infrared galaxies observed with the MIRI/MRS IFU. We perform this calibration and compare the accuracy of six different combinations of wide MIRI filters. We can summarize our results as follows:

\begin{itemize}

\item For the PAH 7.7\,\micron\ feature, the F770W filter traces the intensity of the PAH flux, while F560W provides the best short-wavelength anchor for estimating the continuum. Using Eq.~\ref{eq:fpah77_cal}, both F1500W and F2100W similarly provide good results when either is used as the long-wavelength anchor, yielding PAH flux estimates that are typically accurate within $\sim7\%$.

\item For PAH 11.3\,\micron, the F1130W filter traces the flux of this feature, while F1000W provides the best short-wavelength anchor for estimating the continuum. If only three filters are used (Eq.~\ref{eq:fpah11_cal}), the best filter to use as the long-wavelength anchor is either F1500W, or F2100W. These combinations do not require spectral calibrations for sources with a weak or moderate 9.7\,\micron\ silicate absorption feature ($S_{\mathrm{sil,phot}}>-0.6$), but sources with significant absorption will have significantly underestimated PAH flux up to 25--50\%. Incorporating F560W as a fourth filter (alongside F1000W, F1130W, and either F1500W or F2100W) for PAH 11.3\,\micron\ can be used to account for the strength of the silicate absorption feature and correct underestimated PAH fluxes for obscured sources ($S_{\mathrm{sil,phot}}<-0.6$, see Eq.~\ref{eq:S_sil} and Eq.~\ref{eq:fpah_10umfix}). With this additional filter, the PAH 11.3\,\micron\ flux can be estimated for both absorbed and unabsorbed sources to a typical accuracy of within $\sim5\%$ or $\sim8\%$ if F1500W or F2100W is used, respectively.

\end{itemize}

Using this prescription, large-area maps can be made of the PAH 7.7 and 11.3\,\micron\ flux. These can be used to probe physical conditions of PAH grains like their grain size and ionization distributions, especially when combined with the PAH 3.3\,\micron\ flux extracted from NIRCam photometry.

\section{Acknowledgements} \label{sec:acknowledgements}
This work was supported in part through a Visiting Graduate Researcher Fellowship (VGRF) at Caltech/IPAC. We thank the anonymous referee for their comments that improved this work. G.D. thanks S. Beiler and M. Cushing for helpful conversations about synthetic photometry. H.I. acknowledges support from JSPS KAKENHI grant No. JP21H01129 and the Ito Foundation for Promotion of Science. AMM acknowledges support from NASA ADAP grant \#80NSSC23K0750 and from NSF AAG grant \#2009416 and NSF CAREER grant \#2239807. V.U acknowledges funding support from NSF Astronomy and Astrophysics Grant (AAG) No. AST-2408820, NASA Astrophysics Data Analysis Program (ADAP) grant No. 80NSSC23K0750, NASA Astrophysics Decadal Survey Precursor Science (ADSPS) grant No. 80NSSC25k7477, and STScI grant Nos. HST-AR-17063.005-A, HST-GO-17285.001-710 A, and JWST-GO-01717.001-A.

The JWST data presented in this article were obtained from the Mikulski Archive for Space Telescopes (MAST) at the Space Telescope Science Institute. The specific observations analyzed can be accessed via \dataset[doi: 10.17909/pydq-xz94]{https://doi.org/10.17909/pydq-xz94}.

\software{Astropy \citep{astropy:2013, astropy:2018, astropy:2022}, CAFE (Diaz-Santos et al., in prep.), Matplotlib \citep{Hunter:2007}, NumPy \citep{harris2020array}, SciPy \citep{2020SciPy-NMeth}}

\newpage
\bibliography{citations}{}
\bibliographystyle{aasjournal}

\end{CJK*}
\end{document}